\newcommand{\be}{\begin{equation}}
\newcommand{\ee}{\end{equation}}
\newcommand{\ber}{\begin{eqnarray}}
\newcommand{\eer}{\end{eqnarray}}

\newcommand{\lsim}{\raisebox{-0.7ex}{$\stackrel{\textstyle <}{\sim}$ }}

\documentstyle[preprint,epsfig,epsf,aps,floats]{revtex}
\begin{document}
\tighten
\preprint{\vbox{
\hbox{INT-PUB 02-27}}}
\bigskip
\title{Strangeness Nucleation in Neutron Star Matter}
\author {Travis Norsen}
\address{Dept. of Physics, University of Washington, Seattle, WA 98195}
\date{\today}
\maketitle

\begin{abstract}

We study the transition from npe-type nuclear matter (consisting of
neutrons, protons, and electrons)
to matter containing strangeness, using a Walecka-type model predicting
a first-order kaon-condensate phase transition.
We examine the free energy of droplets of K-matter as the density, 
temperature, and neutrino fraction are varied.  Langer nucleation
rate theory is then used to approximate the rate at which critical droplets
of the new phase are produced by thermal fluctuations, thus giving an
estimate of the time required for the new (mixed) phase to appear at
various densities and various times in the cooling history of the
proto-neutron star.  We also discuss the
famous difficulty of ``simultaneous weak interactions'' which we
connect to the literature on non-topological solitons.  Finally, we 
discuss the implications of our results to several phenomenological 
issues involving neutron star phase transitions.

\end{abstract}

%\pacs{PACS numbers(s): 13.15.+g,13.75.Jz,26.60.+c,97.60.Jd}

\section {Introduction}

A `hot' topic recently in nuclear physics has been the attempt to
understand the production of strangeness in the expanding plasma
thought to be created in relativistic collisions of heavy ions.  There
are several aspects to this complex and difficult problem, including:
how (and, indeed, whether) a 3-flavor deconfined quark-gluon plasma is
produced by the initial collision; how this strange matter equilibrates,
expands, and cools; and how it eventually re-hadronizes, hopefully giving
rise to a unique observable signal.  The goal
of these studies is to better understand the phase structure of QCD in
the low-density, high-temperature region of the phase diagram.

A different but complementary problem involves the study of
nuclear matter at high-density and low temperature, such as that 
existing in neutron star interiors.  Here, matter consisting of 
neutrons, protons, and electrons 
(npe matter) is crushed during the gravitational collapse of the 
parent star.  (Actually, what we call npe matter in this paper may 
also include muons and neutrinos.)  As the star collapses weak 
interactions convert electrons and protons into neutrons, with 
electron neutrinos produced copiously as a by-product.  Additionally, 
neutrinos of all flavors are created through brehmstrahlung-type 
collisions among the warm nucleons.  As the neutrinos diffuse outward 
and radiate away, the proto-neutron star cools from an initial 
temperature of several tens of MeVs, and eventually settles into the 
familiar ground state of a neutron star with central density several 
times nuclear matter density.

It is now understood that at sufficiently high density, a
phase transition will occur in which the neutrons and protons of npe
matter are replaced by deconfined quarks (including strange quarks). 
\cite{cp,alcock,fuller}
A related mechanism for the appearance of strangeness in neutron star
matter is kaon condensation, which may become possible
at densities somewhat lower than the density required for the
deconfining transition described above.  The possibility of kaon 
condensation was first pointed out by Kaplan and Nelson, who were
motivated by the chiral Lagrangian prediction of an attractive 
interaction between $K^-$ and nuclear matter. \cite{KN}

Intuitively, one can understand the phenomenon of kaon condensation in
the following way.  As the density of npe matter is increased, and
assuming that all particle species are in equilibrium with respect
to the weak interactions, the electron chemical potential increases.
Simultaneously, the effective
mass of an in-medium $K^-$ will decrease, due to the attrative interaction
mentioned above.  Therefore, at some critical density, the electron 
chemical potential (i.e., the energy of the electrons at the top of the
Fermi sea) will become greater than the effective mass of the $K^-$.  It
then becomes energetically favorable for the high-energy electrons to
decay according to
\be
e^- \rightarrow \nu_e + K^-
\ee
at which point a condensate of $K^-$ will form.

This picture is somewhat over-simplified, as the production of
kaons does not proceed exclusively by the decay of energetic electrons,
but may also occur via fully-hadronic weak couplings, e.g., 
$n \rightarrow p + K^-$.  Also, the notion of a uniform condensate of kaons
is probably wrong.  As pointed out by  Glendenning \cite{G1}, the 
presence of two separate conserved charges in neutron star matter (namely,
baryon number and electric charge) gives rise to the possibility of a
mixed phase of N-matter
and K-matter existing over a wide range of densities and pressures in the
star.  (By ``N-matter'' we mean matter not containing kaons, but
without the restriction of charge neutrality implied by ``npe matter''.)
This mixed phase will have the form of a Coulomb lattice -- the now
standard ``pasta'' structure consisting of droplets of the new phase
immersed in a background of N-matter (at low densities), with 
rods and slabs replacing droplets as the density is increased, and 
finally with the role of the two phases interchanging in the transition 
toward homogenous K-matter as the density is increased still further.
\cite{GS}

In this paper, we will assume a model which predicts just this kind of
first-order phase transition to kaon-condensed matter.  For the 
description of nuclear matter, we use a Walecka-type model
in which the interactions of strongly-interacting particles are 
mediated by $\sigma$, $\omega$, and $\rho$ mesons, treated in the
mean-field approximation.  Kaons are included in the model on the
same footing as the nucleons, as described in the next section.  

Our goal is, however, not to study the structure of the ground state 
predicted by this model (and its subsequent effect on the global 
neutron star properties such as the mass-radius relation), but rather
to study the {\it nucleation} of kaons,
that is, the process by which the initial proto-neutron star matter
with zero strangeness acquires strangeness through the spontaneous
appearance of
K-matter droplets.  We will begin this study by using the theory
of homogenous nucleation due to Langer. \cite{langer}  Here one
assumes that at a given temperature, there are constant fluctuations
producing small short-lived droplets of the new phase.  Below a certain
critical size, the surface energy cost of these droplets wins out over
the volume term and the droplets shrink away.  For some critical
size, however, the free energy gained from the production of a large
volume of the new (energetically favored) phase is just large enough
to cancel the cost of surface energy, and the droplet will spontaneously
grow.  In
the case studied here, because the two phases have non-zero electric
charge densities, there is also a Coulomb term in the free energy, 
which becomes large for large droplets.  Hence, a super-critical 
droplet will not grow forever, but will reach a stable size at which
the energy gain between the volume and surface energies just balances
the energy cost of the Coulomb energy.

Langer showed that the expected time needed for this kind of phase
transition to proceed depended on the probability of a critical droplet
being produced by thermal fluctuations, and also on the growth rate of
such a critical droplet along the unstable direction in configuration
space.  Assuming a thermal distribution of these
fluctuations, the nucleation rate per unit volume can be written
\be
\Gamma \sim I_0 e^{-\Delta F(R_{crit})/T}
\label{nucrate}
\ee
where $\Delta F(R_{crit})$ is the excess free energy of a critical 
droplet, $T$ is the temperature, and $I_0$ (the ``prefactor'') is a 
microscopic fluctuation rate related to the growth rate of a 
super-critical droplet, the thermal conductivity of the medium, and
other properties.  \cite{langer,linde,fraga}  
The prefactor is often approximated by dimensional analysis as simply
$I_0 \sim T^4$.  We will discuss this approximation 
later.  Our goal is to apply this nucleation rate theory to the case
of cooling proto-neutron star matter in order to better understand
exactly when, where, and how the transition from npe-type matter
to the kaon-condensed phase occurs.  

Our work generally follows previous work on the nucleation of quark matter 
droplets in neutron star matter, which has been studied extensively.
\cite{olesen,heisel,iida,shukla}
The case of kaon condensation is physically unique, however, since
it is the only realistic proposed example of a direct first order
transition 
from npe-type neutron star matter to matter containing strangeness.
In the case of the deconfinement transition mentioned above, for
example, and even at densities, temperatures, and pressures for which
3-flavor quark matter is energetically favored, the transition is
indirect in the following sense:  an intermediate stage of 2-flavor
quark matter will be produced first, with strange quarks then slowly
appearing during a smooth crossover or second-order transition to
the 3-flavor ground state.  This latter transition is, of course, 
slow due to the weakness of the weak interactions.  In this scenario,
however, the original nucleation events which take one from npe
matter to deconfined quark matter, need not involve the weak 
interactions at all.

The case of kaon condensation is radically different, since here
there is no intermediate zero-strangeness state which might allow
for fast nucleation followed by a slow but smooth growth of the
strangeness-containing fields.  Instead, the thermal fluctuations
responsible for nucleation events must directly involve the 
weak-interaction processes which produce kaons.
The difficulties posed by the widely varying timescales involved
in these two sorts of processes (thermal fluctuations and weak
interactions) will form the theme of our discussion.

The outline of the rest of the paper is as follows.  In Section \ref{sec2} 
we present the details of the nuclear mean field theory used for the
subsequent discussion.  In Section \ref{sec3} we use this model to extract 
information about the free energy of K-matter droplets of different
sizes, as the background baryon number density and temperature are
varied.  This allows us to use Eq. (\ref{nucrate}) to estimate the nucleation
rate for the kaon-condensate phase transition.  In Section \ref{sec4} we 
discuss in more detail the underlying mechanism for the fluctuations which
produce the nucleation, and thereby analyze the trustworthiness of the
estimates.  Here we also make contact with the literature on Q-balls
and argue that the problem of direct strangeness
nucleation in neutron star matter may be profitably considered as an
instance of Q-ball nucleation.  Finally, in Section \ref{sec5} 
we summarize the findings, discuss the relevance of our results to 
phenomenological issues in neutron star physics, and indicate some
proposals for future investigation.

\section {Mean-Field Theory Description of Kaon Condensation}
\label{sec2}

In this section we briefly review the model proposed by Glendenning and
Schaffner \cite{GS} which predicts a first-order transition from nuclear
matter to the kaon-condensed phase.  The model begins with a relativistic
Walecka-type Lagrangian describing the neutron and proton fields, as well
as the $\sigma$, $\omega$, and $\rho$ mesons which mediate their 
interactions:
\begin{eqnarray}
{\cal L}_N \!=&& \overline{\Psi}_N \! \left( i\gamma^\mu
\partial_\mu-m_N^\ast
-g_{\omega N}\gamma^\mu V_\mu -g_{r N}\gamma^\mu
\vec{\tau}_N\cdot \vec{R}_\mu \!\right)\! \Psi_N \nonumber
\\
&&{} +\frac{1}{2}\partial_\mu \sigma
\partial^\mu\sigma-\frac{1}{2}m_\sigma^2\sigma^2-U(\sigma)-\frac{1}{4}
V_{\mu\nu}V^{\mu\nu} \nonumber 
\\ 
&&{} +\frac{1}{2}m_\omega^2V_\mu
V^\mu-\frac{1}{4}\vec{R}_{\mu\nu}
\cdot\vec{R}^{\mu\nu}+\frac{1}{2}m_r^2\vec{R}_\mu \cdot
\vec{R}^\mu,
\end{eqnarray}
where $m_N^\ast = m_N-g_{\sigma N}\sigma$ is the nucleon effective mass, which
is reduced compared to the free nucleon mass due to the scalar field $\sigma$.
The vector fields corresponding to the omega and rho mesons are given by
$V_{\mu\nu} = \partial_\mu V_\nu - \partial_\nu V_\mu$, and $ \vec{R}_{\mu\nu}
= \partial_\mu \vec{R}_\nu -\partial_\nu \vec{R}_\mu $ respectively.  
$\Psi_N$ is the nucleon field operator with $\vec{\tau}_N$ the nucleon
isospin operator.

In addition to the usual kinetic, mass, and interaction terms for the
nucleon and meson fields, the model also includes cubic and quartic 
self-interactions of the $\sigma$ field:
\be
U(\sigma) = \frac{1}{3}b m_N (g_{\sigma N} \sigma)^3 + \frac{1}{4} c
(g_{\sigma N} \sigma)^4
\ee
where $b$ and $c$ are dimensionless coupling constants.  These coupling
constants (as well as the three nucleon-meson couplings: $g_{\sigma N}$,
$g_{\omega N}$, and $g_{r N}$) are chosen to reproduce the empirical
properties of nuclear matter at saturation density. \cite{SW,Gbook}

Kaons are included in the model in the same fashion as the nucleons, by
coupling to the $\sigma$, $\omega$ and $\rho$ meson fields. There exist in the
literature several meson-exchange Lagrangians which attempt to describe
kaon-nucleon interactions. A detailed discussion of these models and their
relation to the Chiral Lagrangian description proposed by Kaplan and 
Nelson \cite{KN} can be
found in papers by Pons {\it et al.} \cite{pons} and Prakash {\it et al.}
\cite{kpe,prakash}.  In the present paper we employ the
Lagrangian proposed by Glendenning and Schaffner. \cite{GS}
The kaon Lagrangian is given by
\begin{equation}
{\cal L}_K =({\cal D}_\mu K)^{\dag} ({\cal D}^\mu K)
 - m_K^{\ast 2} K^{\dag} K 
\end{equation}
where $K$ denotes the isospin doublet kaon field.  The covariant derivative $
{\cal D}_\mu = \partial_\mu+ig_{\omega K} V_\mu+ig_{r K} \vec{\tau}_K \cdot
\vec{R}_\mu $ couples the kaon field to the vector mesons and the kaon
effective mass term $m_K^\ast = m_K-g_{\sigma K}\sigma$ describes its coupling
to the scalar meson. $\vec{\tau}_K$ is the kaon isospin operator. 

The vector
coupling constants are determined by isospin and quark counting rules
\cite{GS} and are given by $g_{\omega K}=g_{\omega N}/3$ and $g_{r
K}=g_{r N}$. The scalar coupling is fixed by fitting to an empirically
determined kaon optical potential in nuclear matter.  The real part of
this quantity has been determined from properties of kaonic atoms
to lie in the somewhat wide range
 80 MeV $ \lsim U_K(n_o) \lsim $180 MeV \cite{FGB,RO}.
Here, we choose $U_K(n_o)= 120$ MeV to fix $g_{\sigma K}$.  Lower values 
of the kaon optical potential reduce the strength of the
first-order transition and, eventually, produce instead a second-order
phase transition, while higher values make the first-order transition
stronger.  While the issues discussed in this paper rely on the existence
of a first-order transition, the qualitative conclusions are generally
independent of the specific value of the coupling.  So long as the
transition is first-order, the main effect of changing $U_K(n_o)$ will
be to change the critical density for the onset of the mixed phase,
without severely affecting our discussion of the nucleation properties
near this critical density.

The model as presented so far is a complicated, strongly interacting
field theory which cannot be solved in any reasonable way.  It is therefore
standard to make a mean-field approximation, in which the meson field
operators are replaced by their expectation values.  Because of rotational
invariance only the time-component of the vector fields $V_{\mu}$ and
$\vec{R}_{\mu}$ can have a non-zero expectation value.  Likewise, only
the isospin 3-component of the isovector field $\vec{R}_{\mu}$ can be
non-zero.  The equations of motion for the meson (mean-) fields can be 
simply derived from the above Lagrangians and are given by:
\ber
m^2_\sigma \sigma &=& -\frac{dU}{d\sigma} + g_{\sigma
B}(n^{(s)}_n +n^{(s)}_p) + g_{\sigma k}m_K^* f_K^2 \theta^2  \\ 
m_\omega^2 \omega &=& g_{\omega N} (n_n+n_p) 
- g_{\omega K} f_K^2 \theta^2 (\mu_K+g_{\omega K} \omega 
    + g_{r K} r) \\ 
m^2_r  r &=&  g_{r N} (n_p-n_n) - g_{r K}
f_K^2 \theta^2 (\mu_K+g_{\omega K} \omega + g_{r K} r)
\eer
where the meson fields $\sigma, \omega, r$ now represent the appropriate mean
values.  Here $n_n$ and $n_p$ represent the neutron and proton number
densities, respectively, while $n^{(s)}_n$ and $n^{(s)}_p$ are the 
corresponding scalar densities.  We have substituted $K=(0,K^-)$ and 
$K^- = \frac{1}{\sqrt{2}} f_K \theta e^{-i \mu_K t}$ , where
$f_K$ is the kaon decay constant and $\theta$ is a dimensionless kaon field
strength parameter.  We have neglected to write down the additional small
contributions to the kaon-coupling terms due to finite temperature 
effects, though these are included in our code.  (See Ref. \cite{pons}
for a more explicit presentation of the details of this model for
finite temperature.)  We will be working in the bulk approximation (in which
the meson fields do not vary with position) and so have set to zero the 
gradient terms which would otherwise appear in the equations of motion.

The equation of motion for the kaon field (in terms of $\theta$) is
\be
\label{keom}
0 = \left( ({m_K}^{\ast})^2 - ({\mu_K}^{\ast})^2 \right)\theta
\ee
which indicates that the kaon effective mass $m_K^* = m_K - g_{\sigma K}
\sigma$ and the effective chemical potential $\mu_K^* = \mu_K + X$
must be equal in order for the kaon field to take on a non-zero value.  
Here $X=g_{\omega K} \omega + g_{r K} r$ is the vector field
contribution to the $K^-$ energy.  We also include a Langrangian describing
non-interacting spin-$\frac{1}{2}$ particles to account for the 
presence of electrons, muons, and neutrinos.

The thermodynamic potential per unit volume 
for the nucleon sector is
\ber
\frac{\Omega_N}{V} &=& \frac{1}{2}m^2_{\sigma} \sigma^2 + U(\sigma)
-\frac{1}{2} m^2_{\omega} \omega^2 - \frac{1}{2} m^2_r r^2 \nonumber \\
&& - 2T \sum_{i=n,p} \int \frac{{\mathrm d}^3k}{(2\pi)^3} {\mathrm ln}
\left( 1+ e^{(E(k)-\nu_i)/T} \right)
\eer
where the nucleon energy $E(k) = \sqrt{k^2 + {m_N^*}^2}$.  The
chemical potentials are given by $\mu_p = \nu_p + g_{\omega N} \omega
+ \frac{1}{2}g_{r N} r$ and $\mu_n = \nu_n + g_{\omega N} \omega -
\frac{1}{2}g_{r N} r$.  

The other thermodynamic quantities can be calculated from $\Omega_N$ in
the standard way.  The nucleonic contribution to the pressure, for example,
is $P_N = -\Omega_N / V$, while the number densities and entropy densities
are given by
\ber
n_{n,p} &=& -\frac{\partial \Omega_N/V}{\partial \mu_{n,p}} \nonumber \\
 &=& 2 \int \frac{{\mathrm d}^3k}{(2\pi)^3}f_{F}(E(k)-\nu_{n,p}) \\ 
& & \nonumber \\
s_{n,p} &=& -\frac{\partial \Omega_N/V}{\partial T} \\
\eer
where $f_F(\epsilon) = \left( e^{-\epsilon/T}+1 \right)^{-1}$ is the
Fermi-Dirac distribution function.  The neutron and proton scalar 
densities which enter in the equation of motion for the $\sigma$ field 
are given by
\be
n_{n,p}^{(s)} = 2 \int \frac{d^3k}{(2\pi)^3}\frac{m_N^*}{E(k)}
 f_{F}(E(k)-\nu_{n,p})
\ee
The energy density is determined through the usual relation 
$T s_N = \epsilon_N + P_N - \sum_i \mu_i n_i$ to be
\ber
\epsilon_N &=& \frac{1}{2}m_{\sigma}^2 \sigma^2 + U(\sigma) - \frac{1}{2}
m^2_{\omega} \omega^2 - \frac{1}{2} m^2_r r^2 \nonumber \\
&& + 2\sum_{i=n,p} \int \frac{{\mathrm d}^3k}{(2\pi)^3} E(k) f_{F}
(E(k)-\nu_i) \\
&+& \sum_{i=n,p} n_i(\mu_i - \nu_i).\nonumber
\eer

The thermodynamic potential for the lepton species present (electrons,
muons, and neutrinos) is given by
\be
\frac{\Omega_L}{V} = -\sum_l T g_l \int \frac{{\mathrm d}^3k}{(2\pi)^3}
\left[ {\mathrm ln} \left( 1+e^{-(E_l(k)-\mu_l)/T} \right) \\
  + {\mathrm ln} \left( 1+e^{-(E_l(k)+\mu_l)/T} \right) \right]
\label{lepton}
\ee
where $\mu_l$ denotes the chemical potential for lepton species $l$, and
$g_l$ are the spin-degeneracies:  $g=2$ for electrons and muons, $g=1$
for neutrinos.  $\beta$-equilibrium requires the following constraints
on the chemical potentials:
\be
\mu_K=\mu_e-\mu_{\nu_e} = \mu_n - \mu_p = \mu_{\mu} - \mu_{\nu_{\mu}}
\label{beq}
\ee
The lepton contributions to the pressure, energy density,
and entropy are determined from Eq.(\ref{lepton}) in the usual way.

Finally, the thermodynamic potential for the kaons is given by
\ber
\frac{\Omega_K}{V} &=& \frac{1}{2}f_K^2 \theta^2 \left[ (m_K^*)^2 -
 (\mu_K + X)^2 \right] \nonumber \\
&& + \; T\int \frac{d^3p}{(2\pi)^3}{\mathrm ln}
\left( 1-e^{-(\omega^-(p) - \mu_K)/T}\right)
\eer
where $\omega^-(p) = \sqrt{p^2 + (m_K^*)^2} - X$ is the in-medium
energy of a $K^-$ with momentum $p$.  Again, the kaon contribution to 
the other thermodynamic 
quantities can be found by differentiation.  For a detailed derivation
and presentation of the thermodynamics of this model, see Ref. 
\cite{pons}.  For discussions of similar studies involving mixed-phases
in mean-field-theories in several different physical contexts see, 
e.g., \cite{lamb,RPW,HPS,Loren}.

Our goal here is to study the rate at which K-matter nucleates
in a background of npe-matter at various temperatures and densities.
As discussed previously, this involves calculating the free energy of
droplets of various sizes, in particular, the free energy of the critical
droplet configuration.  The first step toward this end is to produce a 
description of a droplet of K-matter of arbitrary radius.  This involves 
solving the meson field equations of motion subject to various constraints.  

Generally, one solves simultaneously two versions of the meson field 
equations above:  one with $\theta = 0$ describing the N-matter component
of the mixed phase, the other with $\theta \not= 0$ (with the specific
value of $\theta$ determined self-consistently through Eq. (\ref{keom})) 
describing the K-matter.  One requires chemical equilibrium between the
two phases, i.e., that the relevant chemical potentials in the two phases
match.  Additionally, the requirement of overall electric charge neutrality
means that the electric charge densities in the two phases must be of
opposite sign.  One then uses the respective charge densities to calculate
$\alpha = -q^{(N)}/(q^{(K)}-q^{(N)})$, the volume fraction of the 
K-matter phase.  (Here $q^{(N)}$ and $q^{(K)}$ are the charge densities 
of the K-matter and N-matter,
respectively.)  Because we are concerned with droplets of small radius, 
it is also crucial to include correct mechanical equilibrium between 
the two phases.  In Ref.\cite{norsen} this requirement was shown to 
affect rather dramatically the bulk properties of the two phases.  
(See also \cite{Gfinite}.)  For a spherical droplet of K-matter 
in a background of N-matter, this constraint reads:
\be
P_K - P_N = 2\sigma / R
\label{pressure}
\ee
where $\sigma$ is the surface tension between
the two phases, and $R$ is the radius of the droplet.  

Finally, it should be noted that we work at constant baryon number density.
That is, in constructing a sequence of droplets of varying radii, we 
require that the overall baryon density $n_B = \alpha n_B^{(K)} + 
(1-\alpha) n_B^{(N)}$ be held fixed at some specified value.  Here 
$n_B^{(N,K)}$ are the baryon number densities of the N- and K-phases.

Once the equations of motion are solved self-consistently subject to 
these constraints, it is possible to calculate the free energy of a 
given droplet.  The total bulk energy density of a given droplet 
configuration is defined analogously to the overall baryon number 
density:
\be
\epsilon_{bulk} = \alpha \epsilon^{(K)} + (1-\alpha) \epsilon^{(N)}
\label{eden}
\ee
where the energy density in each phase includes the contributions from the
nucleons, leptons, and kaons, as appropriate.  The total bulk energy is 
then found by multiplying this energy density by
the total volume of a Wigner-Seitz cell (defined as the spherical region
including one droplet and containing zero total electric charge), 
$V_{WS}$.  This energy is then supplemented by the surface and Coulomb
contributions:
\be
E(R) = \epsilon_{bulk} V_{WS}(R) + 4\pi R^2 \sigma + E_{Coul}
\ee
where $E_{Coul}$ is found by integrating the electric field energy density
$\epsilon_{Coul}(r) = \frac{1}{2}|E(r)|^2$ throughout the Wigner-Seitz cell.
(Here $|E(r)|$ is the magnitude of the electric field determined by 
Gauss' Law.)   The total entropy of a droplet configuration is found 
analogously, by volume-fraction-weighted averaging of the individual 
phase entropy densities, then multiplying by the volume of the 
Wigner-Seitz cell.  Thus, the total free energy of a droplet can be 
calculated as:
\be
F(R) = E(R) - T S(R).
\ee
Only the R-dependence has been indicated explicitly, but, of course, 
the energy and entropy both depend strongly on the fixed baryon number
density, the temperature, neutrino fraction, {\it etc}.  

\begin{figure}[t]
\centering
{
\epsfig{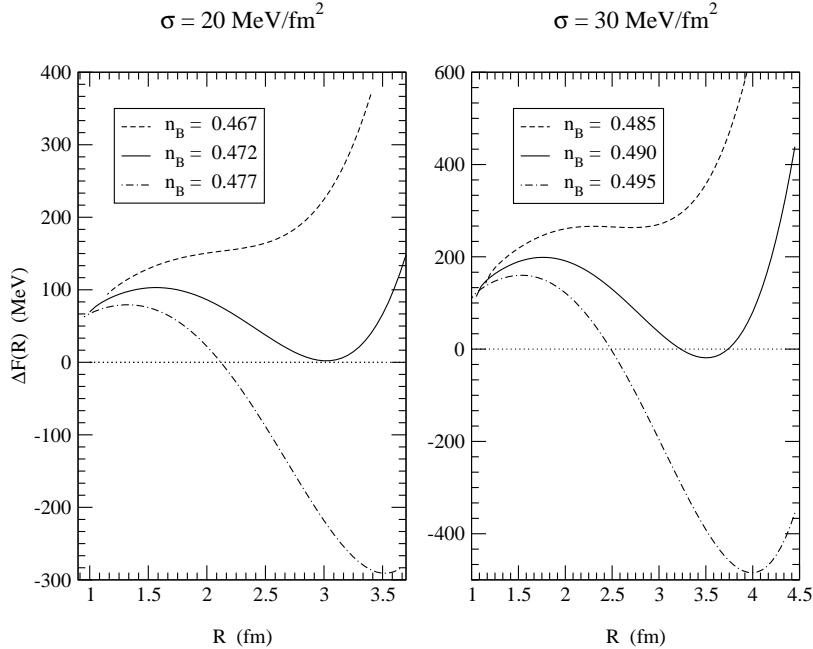}
}
\caption{Free Energy cost of kaon droplets (as a function
of the radius) for two different values of the surface tension, $\sigma$.  
Each plot is shown near the critical density for the transition (which is
slightly different for the two values of $\sigma$), with additional curves
drawn just above and just below the critical density, showing the development
of an energetically favored droplet structure as the density is increased
past the critical density.}
\label{fig1}
\end{figure}

We are interested, however, not merely in the free energy of a given
configuration, but, rather, the change in free energy required to
produce various droplets.  Hence, we also solve for pure, charge neutral
npe matter at a given baryon density and calculate its free
energy density, $f_{npe} = \epsilon - T s$.  We are then led to define
\be
\Delta F(R) = F(R) - f_{npe} V_{WS}.
\ee
which represents the free energy cost of a transition from electrically
neutral npe matter (the initial state of an evolving proto-neutron star)
to a single droplet of kaon-condensed matter of radius $R$. Fig. \ref{fig1} 
shows $\Delta F(R)$ as a function of
$R$ near the critical density for the transition, for two different values
of the surface tension, $\sigma = 20 MeV / fm^2$ and $\sigma = 30 MeV /
fm^2$.  

We have shown the free energy curves for two values of $\sigma$ in order
to illustrate the role this quantity plays.  As expected, smaller values
of $\sigma$ reduce both the size and free energy cost of a critical droplet.
In what follows, we will simply pick the value $\sigma = 30 
{\mathrm MeV}/fm^2$,
a value suggested by Glendenning's study of the boundary between the
two phases \cite{Gsurf}, as well as by our own earlier work on this
model \cite{norsen}.

In the following section we will use this type of curve to acquire 
information about the free energy cost of a critical droplet under various
conditions of temperature, density, and neutrino fraction.

\section {Droplet Free Energy and Nucleation Rates}
\label{sec3}

As mentioned in the introduction, one can understand the production of
kaons in the neutron star matter as being due to the weak decay of an
electron (or, equivalently, the change of a neutron into a proton plus
a kaon).  By definition, above the critical density this transition is
energetically favorable, i.e., exothermic.  So we may schematically
write
\be
e^- \rightleftharpoons \nu_e + K^- + {\mathrm Heat}.
\label{equil}
\ee
The excess heat generated by the production of kaons will eventually
be radiated away in the form of photons and neutrinos as the neutron
star cools, but that is not our interest here.  Rather, our goal is
to understand how the free energy of droplet configurations depends on
the temperature.  Qualitatively, one can guess the correct answer by
applying LeChatelier's principle to the equilibrium indicated in
Eq.(\ref{equil}):  raising the temperature
will tend to push the equilibrium back toward the left.  That is, raising
the temperature should lower the energetic favorability of kaon droplets.  

\begin{figure}[t]
\centering
{
\epsfig{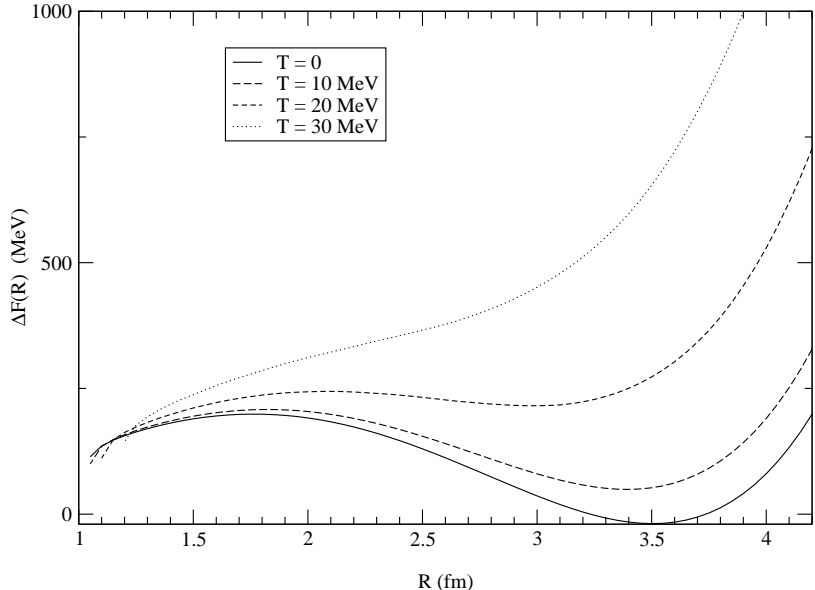}
}
\caption{Free energy of droplets (relative to neutral npe matter with the
same baryon number density) as a function of radius at temperatures 
varying between zero and 30 MeV.  As expected, increasing the temperature
decreases the energetic favorability of the kaon phase.}
\label{fig2}
\end{figure}

This prediction is borne out by our calculations, as illustrated in Fig.
\ref{fig2}.  As the temperature is increased from zero, the free energy
of kaon droplet configurations increases relative to that of neutral npe
matter at the same density and temperature.  This increase (which is as
large as several hundred MeV for typical droplets) means that, at a 
density where the mixed
phase would be the ground state at $T=0$, the mixed phase is no longer
favored at higher temperature.  In other words, turning up the temperature
increases (slightly) the critical density for the transition.

\begin{figure}[t]
\centering
{
\epsfig{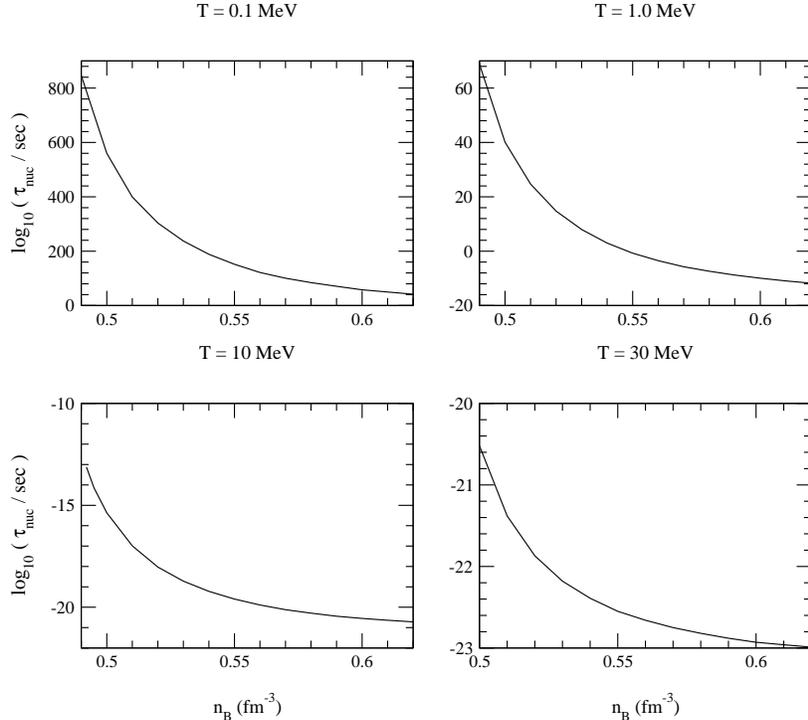}
}
\caption{Expected time for the nucleation of a single K-matter droplet
in a typical Wigner-Seitz volume.  Shown is the ${\mathrm log}_{10}$ of
the nucleation time in seconds as a function of baryon number density,
at several different temperatures.}
\label{fig3}
\end{figure}

However, because the kaon phase droplets must be produced by thermal
fluctuations, we expect the thermal nucleation rate of the kaon phase to 
be much larger at higher temperatures.  In order to estimate the speed 
with which droplets of the kaon phase are produced, we use the Langer 
nucleation rate theory discussed in the introduction.  The nucleation 
rate per unit volume is estimated as:
\be
\Gamma \sim T^4 e^{-\Delta F(R_{crit})/T}
\label{langer}
\ee
where T is the temperature and $R_{crit}$ is the critical droplet radius
(at a given temperature and baryon density).  The free energy cost of this
critical droplet configuration (i.e., the height of the barrier which must
be crossed to produce a stable droplet of the new phase) can be easily 
read off of graphs like the ones already shown.  One may then plug 
directly into Eq. (\ref{langer}) to give the nucleation rate per unit 
volume.  In order to convert this rate into an intuitively meaningful 
quantity, we calculate the expected time for a single nucleation event 
in a single typical Wigner-Seitz cell of volume $V_{WS} = 10^3 fm^3$.  
This time is given by
\be
\tau = \frac{1}{\Gamma V_{WS}} = \frac{e^{\Delta F(R_{crit})/T}}
{V_{WS}T^4}
\ee
In Fig. \ref{fig3} we show this nucleation time as a function of density
for several different temperatures.  (Note the log scale!)  As expected,
the nucleation time is a very strong function of the temperature.  At a
temperature of $0.1$ MeV, the expected nucleation time is many, many times
longer than the age of the universe across the entire density range of
the mixed phase.  At $T=1.0$ MeV the nucleation time is prohibitively
long at the lower end of the mixed phase density regime, but is less 
than one second at densities above about $n_B \sim 0.55 fm^{-3}$, which
is something like $\frac{1}{2}n_o$ above the critical density.  At
higher temperatures, the nucleation proceeds almost immediately across
the entire density range.  

Naively, this leads to the conclusion that the neutron star settles into
its ground state without any delay as it cools down from an initial 
temperature of several tens of MeV.  In the initial, hot conditions,
thermal fluctuations are sufficiently fast and sufficiently numerous to
seed K-matter droplets wherever those droplets are energetically favored.
The sizes and distances between adjacent kaon structures may undergo
tiny changes as the matter cools, and perhaps the outer edge of the 
mixed phase extends outward somewhat as the critical density decreases,
but generally the neutron star includes the full mixed phase from birth.

This picture is complicated by at least two factors which have not
been discussed explicitly until now.  The first of these is the presence,
in the early stages of proto-neutron star cooling, of a significant
neutrino fraction.  The second question is whether or not we should
believe the rate estimates just given, since the nucleation rate 
prefactor (estimated above as $T^4$) describes the rate of 
microscopic fluctuations, which, in the present case, consist of
{\it weak} interaction processes.  These processes are, after all,
weak, so one might doubt that the naive estimate based simply on
dimensional analysis is appropriate.  This issue will be discussed
in a subsequent section; for now, we will turn to the question of
the effects of neutrinos on the nucleation of K-matter.  (For a
related discussion see Ref.\cite{pmpl})

\begin{figure}[t]
\centering
{
\epsfig{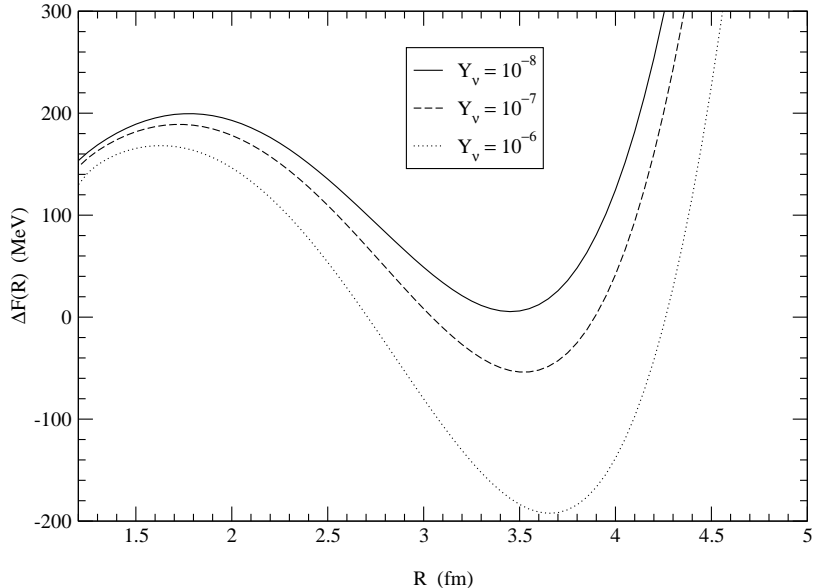}
}
\caption{Free energy of droplets (relative to neutral npe matter with the
same baryon number density) as a function of radius for various small
neutrino densities.}
\label{fig4}
\end{figure}

Returning to Eq.(\ref{equil}) and again applying LeChetalier's principle,
we guess that the presence of a non-zero density of electron neutrinos
will (like high temperature) suppress the appearance of K-matter.  This
guess turns out to be correct, but for slightly complicated reasons. 
Turning on a non-zero electron neutrino fraction
\be
Y_{\nu_e} = \frac{n_{\nu_e}}{n_B}
\ee
forces the chemical
potentials for electrons, kaons, and the baryons to adjust according to
the constraint of constant $n_B$ and $\beta$-equilibrium, Eq. 
(\ref{beq}).  Surprisingly, doing this actually lowers the free
energy cost of kaon droplets, as indicated in Fig. \ref{fig4}.

The reason for this can be understood as follows.  The main effect of
turning on a chemical potential for neutrinos, is to increase as well
the chemical potential for electrons since the constraint of fixed
baryon number density doesn't allow $\mu_n$ and $\mu_p$ much freedom
to adjust.  But increasing the number density of electrons decreases
the (positive) electric charge density of the N-matter outside a kaon
droplet.  Hence, more of this matter is needed to cancel the negative
charge of the droplet itself, and the volume fraction of K-matter, 
$\alpha$, decreases.  But the overall free energy density depends on
$\alpha$ through Eq.(\ref{eden}).  Finally, since the energy density of
K-matter is somewhat larger than that for electrically neutral npe
matter (due not only to the presence of kaons, but also to the higher
local baryon density) while that of the surrounding N-matter is somewhat 
lower than neutral npe matter (due to the correspondingly lower baryon 
number density here), the overall energy density of the mixed phase is 
actually decreased by the decrease in $\alpha$ coming from the non-zero 
neutrino density.

However, because of the requirement of
overall electric charge neutrality and the fact that the K-matter is 
always negatively charged, the N-matter component outside of the droplet
must have a positive electric charge density.  As
we have just seen, the presence of even a small non-zero chemical potential 
for electron neutrinos also increases the density of electrons, and thereby
reduces the positive charge in this region.  As expected, a larger neutrino
fraction will decrease the electric charge density here even more, and
eventually a point is reached at which the charge density ceases to be
positive.  Then it is no longer possible to define a Wigner-Seitz cell, 
{\it i.e.}, no longer possible to satisfy the requirement of electric 
charge neutrality.  

Actually, this point is reached for very moderate values of the neutrino
fraction, especially at low densities (near the critical density for the
transition) where, by definition, the electric charge density in the
N-matter phase is positive, but small.  We find that the pressure 
equilibrium condition Eq.(\ref{pressure}) also has a comparable effect 
on the electric charge densities via the electron chemical potential.
(This issue was discussed in Ref.\cite{norsen}.)  In particular,
larger structures at a given baryon density have lower electron chemical
potentials; hence it is possible to elude the effect of non-zero $Y_{\nu}$
by producing larger K-matter droplets.  Thus, at a given $Y_{\nu}$ and
a given baryon density, there will be a minimum size $R_{min}$ consistent
with charge neutrality, and such structures are therefore not allowed.  
Electron Debye screening will also play a role in forbidding 
structures with large radii, so the region of 
$n_B - Y_{\nu}$ parameter space in which no mixed phase can exist is 
even somewhat larger than suggested here. \cite{norsen}

\begin{figure}[t]
\centering
{
\epsfig{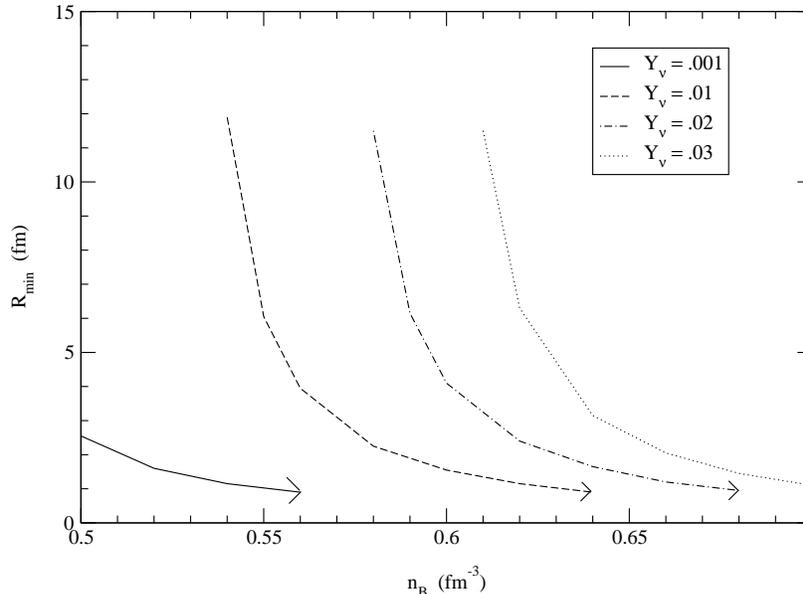}
}
\caption{Minimum droplet radius consistent with overall electric
charge neutrality, as a function of baryon number density, for 
several different values of the neutrino fraction $Y_{\nu}$.  Here
the temperature is fixed at $T=0$.}
\label{fig5}
\end{figure}

We show in Fig.\ref{fig5} this minimum radius as a function of baryon
density, for several different values of the electron neutrino fraction,
$Y_{\nu}$.  At a given $Y_{\nu}$, the minimum radius allowed by global
charge neutrality begins to diverge as one comes down in density.  The
density at which this quantity diverges (or, if we were to consider also
the effects of Debye screening, the density at which this quantity exceeds
5-10 fm) acts as an effective critical density for the onset of the mixed
phase.  Hence, early in the evolution of the proto-neutron star (PNS), 
when the neutrino fraction is (at least) a few percent, no mixed phase 
will be 
produced at densities less than $0.60 - 0.65 fm^{-3}$.  This is to be 
compared to the critical density at zero temperature and zero neutrino 
fraction, $0.49 fm^{-3}$.  Actually, the electron neutrino fraction in 
the first seconds of PNS evolution may be closer to ten percent, in
which case the effective critical density below which no stable kaon
mixed phases exists will be pushed upward to around $0.80 fm^{-3}$,
several factors of $n_0$ above the nominal ($T=0$, $Y_\nu = 0$)
critical density.  The upshot is that, in the neutrino-rich conditions of
the early PNS evolution, the critical density for the onset of the kaon
mixed phase is increased significantly compared to the nominal critical
density.

This obviously modifies the naive inference from Fig.\ref{fig3}
that the mixed phase will be produced quickly and easily during the
early, hot part of the PNS evolution.  Because the high temperatures
occur at a time of high neutrino density, there is a wide range of
densities over which no mixed phase can be formed at these early
times.  Hence, due to the suppression of K-matter by neutrinos, a 
PNS with central density not too far above the nominal
critical density for the onset of the mixed phase may fail to
produce any K-matter during its cooling, until the temperature is 
so low that nucleation can not occur in reasonable times.  This
suggests a scenario in which the neutron star could exist 
indefinitely in a metastable state.  However, further analysis
is required to support this hypothesis -- in 
particular, we must address the question of the reliability of the
nucleation rates in Fig.\ref{fig3} given the problem of simultaneous
weak interactions. \cite{alcock}

\section {Fluctuations and the Problem of Simultaneous Weak Interactions}
\label{sec4}

As a first guess, one might suppose that the nucleation of K-matter 
droplets is driven by thermal fluctuations in the local number density
of baryons.  At a sufficiently high baryon number density, the phase
transition to the kaon condensed phase becomes second order, and the
kaon field may grow smoothly without having to overcome an energy
barrier.  Hence, one might suppose that the first-order transition
could occur by this same mechanism operating locally:  a thermal
fluctuation produces a baryon overdensity in a local region of radius
$1-2 fm$.  The kaon field then spontaneously ``fills in'' this
region, producing a stable droplet of K-matter
which could then grow to the stable size.  In this picture, the nucleation 
rate would be governed by the frequency of sufficiently over-dense and 
sufficiently large baryon number fluctuations.

However, there is an immediate and fatal problem with this basic picture.
Sufficient fluctuations in baryon density are no-doubt plentiful.  A
typical $1-2 fm$ region contains of order ten baryons, so assuming 
simple $\sqrt{N}$ fluctuations, a factor of two overdensity is only
three standard deviations away from the mean.  But the lifetime
of such a fluctuation is limited to strong interaction timescales.  Indeed,
in nuclear matter at these densities, the speed of sound is a sizable
fraction of
the speed of light, $c$.  Hence, the lifetime of a baryon number density
fluctuation is of order $\tau \sim R/c \sim 10^{-23} sec$.  As is 
well-known from studies of a second-order kaon condensate phase
transition, however, the time-scales needed for the development of an
appreciable value of the kaon field are $10-15$ orders of magnitude longer
than this. \cite{muto} Similar conclusions are found for the appearance 
of strange quark matter from an initial state of 2-flavor quark matter
in neutron stars. \cite{dai,ghosh} 
That is, not surprisingly, the time-scale for the development
of the kaon field is typical of the weak interactions. 

The production of K-matter droplets, therefore, cannot be driven by 
density fluctuations in the background of baryons.  The microscopic
kaon-production rate is simply too slow to keep up with the strong
interaction timescales that govern such fluctuations, so long before
any appreciable value for the kaon field in the fluctuation has
developed, the fluctuation evaporates.  

\begin{figure}[t]
\centering
{
\epsfig{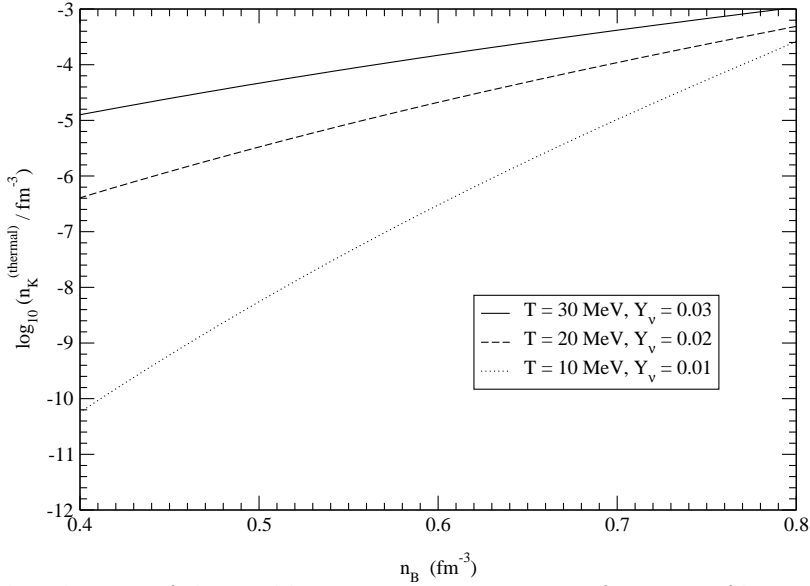}
}
\caption{Number density of thermal kaons in npe-matter as a function
of baryon number density.  The three curves correspond (very) 
approximately to three times during the cooling history of the 
proto-neutron star.  As the star cools, the density of thermal
kaons drops substantially and is effectively zero in the context of 
the probability of seed-production considered below.}
\label{fig6}
\end{figure}

We conclude that the nucleation rate will be goverened by thermal fluctuations
producing fluctuations in the local kaon number density itself.  Moreover,
so long as the timescale for the spontaneous growth of a critical
droplet is small compared to weak interaction timescales, this rate can 
depend only on the thermal number density of kaons,
\be
n_K^{(thermal)} = \int \frac{d^3p}{(2\pi)^3}f_B(\omega^-(p) - \mu_{K})
\ee
where $f_B$ is the Bose-Einstein occupation probability and
$\omega^-(p)$ was defined in Sect.\ref{sec2}.  We have shown in 
Fig.\ref{fig6} this thermal kaon number density for several values of 
the temperature and neutrino fraction in npe-type neutron star matter 
as a function of density.

It is then simple to estimate the probability of a sufficient ``seed''
of kaons appearing.  This calculation is reminiscent of the standard
undergraduate statistical mechanics problem of calculating the
probability that all of the atoms in a box of gas are found to
be in a certain small region of the box.  Here, we must calculate the
probability that a certain number $N$ of the kaons in a kaon gas of
density $n_K^{(thermal)}$ spontaneously appear in some small region
of space with volume $V$.  

For definiteness, consider a large box of volume $V_0$ and total
kaon number $N_0$, with $N_0/V_0 = n_K^{(thermal)}$.  If we treat the 
kaons as classical particles, then the probability of $N$ kaons being 
found in a small
sub-region $V$ of $V_0$ is given by the binomial theorem:
\be
P(N) = \frac{N_0!}{N!(N_0-N)!} \left( \frac{V}{V_0}\right)^N
\left( 1 - \frac{V}{V_0} \right)^{(N_0 - N)}
\ee
Assuming $V \ll V_0$ and $N \ll N_0$, we may approximate
$N_0! \sim (N_0 - N)! N_0^N$ and neglect $N$ in the exponent $(N_0-N)$.
This gives
\be
P(N) \sim \frac{1}{N!}\left( \frac{N_0 V}{V_0} \right)^N
\left( 1 - \frac{V}{V_0}\right)^{N_0}.
\ee
But $N_0 V / V_0 = <N>$, where $<N> = V n_K^{(thermal)}$ is the average
number of thermal kaons in the volume $V$.  Using $<N>/V = N_0/V_0$ we 
have finally
\ber
P(N) &\sim&\frac{<N>^N}{N!} \left(1-\frac{<N>}{N_0}\right)^{N_0} \nonumber
\\
&\sim& \frac{<N>^N}{N!} e^{-<N>}.
\label{poisson}
\eer
(In practice, $n_K^{(thermal)} \ll 1 fm^{-3}$ so the exponential in 
the last line above can be ignored.)

\begin{figure}[t]
\centering
{
\epsfig{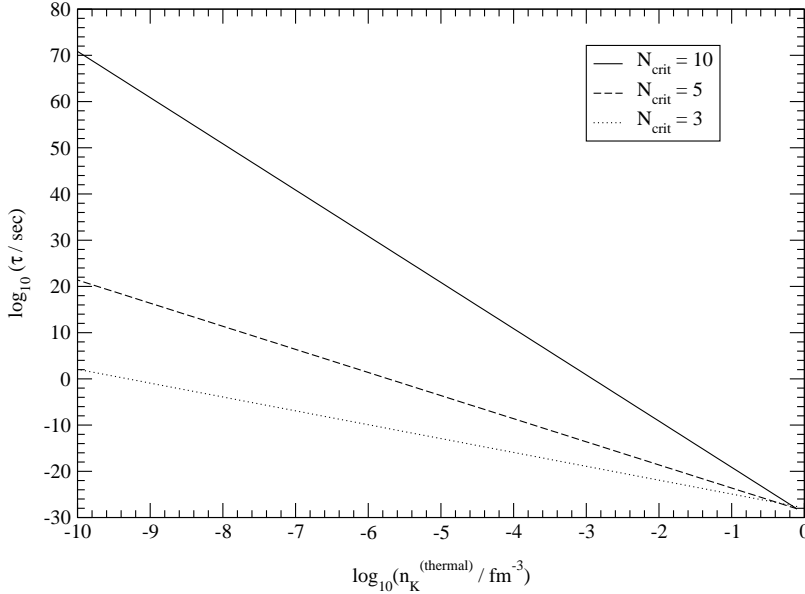}
}
\caption{Expected time for a critical droplet of N kaons to form out
of a spontaneous density fluctuation of kaons with average density
$n_K^{(thermal)}$.}
\label{fig7}
\end{figure}

The number of kaons in a critical droplet varies somewhat with
density.  Near the critical density (at zero temperature and zero neutrino
fraction) the radius of a critical droplet is a few $fm$, for a critical
kaon number of order 100.  At higher densities, the critical radius drops
to only $1-2 fm$, and the corresponding critical kaon number drops to 
only a few.  In Fig. \ref{fig7} we plot the time needed to form such a
critical seed (for various $N$) as a function of thermal kaon number
density.  To estimate the seed-production time from the probability 
considered above, we multiply each probability by the maximum possible
number 
of discrete seeding locations within a typical Wigner-Seitz cell.  This
is of order $V_{WS} n_B \sim 1000 fm^3 \cdot .5/fm^3 \sim 500$.  We then
assume that the arrangement of thermal kaons is ``reshuffled'' on the
fastest possible time-scale, say $\tau \sim 10^{-23} sec$.  (Note that
this assumption is extremely generous.  The actual reshuffling time must
be at least several orders of magnitude greater than this, but here a
generous upper limit on the seed-production rate will suffice.)  Then the
typical time needed for a seed of $N$ kaons to be produced by this
method is given by
\be
\tau(N) \sim \frac{10^{-23}sec}{500\cdot P(N)}
\ee
with P(N) given as a function of $n_K^{(thermal)}$ above.  Actually, 
we should replace $P(N)$ here with $\sum_{n\geq N} P(n)$ but in
practice each term in the sum is negligible compared to the one
previous, so the difference will not affect the result.

We see that the number $N_{critical}$ is indeed critical to the
determination of the rate.  If ten (or more) kaons are needed to produce
a critical droplet, the probability of a sufficient number all showing
up in the same place at the same time is extremely small, and one must
wait prohibitively long for a seed to ever be produced, especially at
lower temperatures where $n_K^{(thermal)}$ is extremely small.  
At higher densities where $N_{crit}$ is lower, of 
order 3-5, the probability is much higher the time needed for a
seed to be produced by random reshuffling may be of order seconds or less
for reasonable values of the temperature.

Comparing Fig. \ref{fig6} and Fig. \ref{fig7}, we see that the time
needed for a seeding event during the initial hot proto-neutron star
configuration ($T=30 MeV$, $Y_\nu = 0.03$) is on the order of a few
seconds or less if $N_{crit} \le 5-10$.  However, as we see from Fig.
\ref{fig5}, electric charge neutrality forbids droplets smaller than
$\sim 2 fm$ for densities below $\sim 0.65 fm^{-3}$.  This is well
above the ($T=0$, $Y_\nu = 0$) critical density, and it is likely 
that such a high density is not reached in the proto-neutron star core,
especially during the early evolution when the star is still hot and
relatively rarefied.

After a few tens of seconds, the PNS temperature drops to 
$T=10$ MeV with the neutrino fraction also dropping to, say, 
$Y_\nu = 0.01$.  Looking again at Fig. \ref{fig6}, we see that the
number density of thermal kaons has dropped dramatically due to
the lower temperature.  The electric charge neutrality constraint
is relaxed somewhat, with droplets of radius $1-2 fm$ now being
allowed at densities below about $n_B \sim 0.58 fm^{-3}$.  There is
a possibility, considering Fig. \ref{fig7}, that seeding may occur
with reasonable speed for densities above this value, where 
$N_{crit}$ is of order $3-5$.  But again, there is a fairly rigid
effective critical density below which nucleation cannot occur.  As
the star continues to cool, the number density of thermal kaons 
drops dramatically, and the seeding time becomes prohibitively long
at all but the very highest densities, $5-7$ times nuclear matter
density, reasonably thought to exist in the neutron star core.

The crude estimates made here for seed production in a homogenous but
fluctuating background of thermal kaons are similar to the estimates in 
\cite{griest,metaxas} of the growth process for sub-critical
Q-balls in a scalar field with a conserved charge.  These authors
assume a thermal distribution of sub-critical Q-balls with
relatively large N, and then consider the rate at which additional
charge is accreted to (and released by) the Q-ball via a random-walk 
process in order to estimate the rate at which critical droplets are
produced.  In our case, the number of kaons in a critical droplet is
not much greater than unity, so we use the Poisson statistics of Eq.
(\ref{poisson}) rather than a Gaussian distribution with $\sqrt{N}$ 
fluctuations.

Actually, there is an exact analogy between the physics of kaon 
nucleation and the problem of Q-ball nucleation, {\it 
i.e.}, the decay of metastable field configurations in the presence of
a conserved charge.  
\cite{coleman,benson,lee1,lee2}
The effective theory for the kaon field in our model will be precisely
the theory of a complex scalar with an effective potential supporting
non-topological solitons, i.e., stable droplets of K-matter.  In 
the limit of infinitely slow weak interactions, this effective theory
will contain a global $U(1)$ flavor symmetry corresponding to the 
conservation
of strangeness.  In this limit, the seeding of K-matter droplets will
occur via a process formally identical to that described in Ref.
\cite{lee2}, namely, a small uniform initial charge density 
(in our case, the background of thermal kaons) producing the
required ``bounce'' configuration by flowing toward a seeding
center.  As discussed in \cite{lee2}, the action associated with
this configuration will be significantly greater than the 
corresponding case for a real scalar which carries no conserved
charge.  The typical nucleation rates in the case of a 
charged scalar will thus be exponentially slower, due to the need
for a global redistribution of charge.

The two estimates of nucleation rates in the present paper can be
thought of as limits in the cases of infinitely slow and infinitely
fast weak interactions.  In reality, of course, the weak interactions
proceed at some intermediate speed.  In the effective theory of
the kaon field described above, inclusion of strangeness-changing
weak reactions will correspond to the addition of
a small symmetry-breaking term in the effective potential, e.g.,
\be
U_{eff}(|K|) \rightarrow U_{eff}(|K|) + \lambda K
\ee
where $\lambda$ is a strongly temperature-dependent constant that
characterizes the rate at which kaons can be produced/absorbed by
background scattering processes involving nucleons and leptons.

The actual rate of K-matter nucleation will depend on both types of
symmetry breaking, that is (1) the initial background thermal kaon
density and (2) the dynamical production of new kaons, though it
is not obvious which mechanism will dominate the 
nucleation for physically relevant temperatures.  In order to treat
this problem reliably, one must construct realistic bounce 
configurations in the kaon effective theory as the initial background
density and dynamical breaking parameter $\lambda$ are varied with
temperature, impose imaginary-time periodicity corresponding to the
(inverse) temperature, and calculate the action of the bounce.  
Qualitatively, one expects the resulting nucleation times to be
intermediate between the fast-weak-interactions and no-weak-interactions
limits in the current paper.  At very least, one should expect that the
slowness of the weak interactions should modify the nucleation rate
prefactor $I_0$ to a more realistic dimensional analysis estimate
\be
I_0 \rightarrow G_F^2 T^8 \sim \frac{T^8}{M_W^4}
\ee
where $M_W$ is the mass of the weak gauge bosons.  At a temperature of
$10$ MeV, this estimate would reduce the expected nucleation time by
a factor of approximately $10^{-16}$ from the times shown in Fig.
\ref{fig3}.  As mentioned above, the issue of nearly-conserved
strangeness will also affect the exponential factor in the rate
equation, thus suppressing the rate even further and perhaps bringing the
results closer to the estimates in Fig. \ref{fig7}.  Only the reliable
calculation outlined above (and currently underway by the present
author) will answer this issue with any certainty,
however.  

\section{Discussion and Conclusions}
\label{sec5}

We have, then, an intriguing possible scenario in which the PNS 
manages to settle into its ground state without forming the 
kaon-condensate mixed phase which is the true ground state
of the system.  At high initial temperatures, the nucleation of
K-matter droplets is suppressed by the presence of neutrinos, even
though at these high temperatures the seeding of droplets is in
principle fast.  Over a wide range of densities kaon droplets
are produced copiously by thermal fluctuations, but they are not 
yet stable due to the increase in effective critical density.  As
the star cools, the restriction coming from the presence of neutrinos
is relaxed, but the intrinsic fluctuation rate drops.  

For an initial PNS core density that is not too far above the nominal
($T=0$) critical density for the formation of a K-matter mixed phase,
therefore, it is likely that the star will cool not into the true
ground state, but, rather, into a meta-stable state consisting of
electrically neutral npe matter.  This scenario may be relevant to
understanding various phenomenological issues.

For example, the apparent existence of anomalously heavy neutron stars
with masses $M \sim 2 M_{sun}$ \cite{zhang,kerk,orosz}
might be explained by the anomalously
stiff equation of state of npe-type matter relative to matter with a
kaon condensate.  Generally, if the various possible phase
transitions thought to occur in neutron star matter can be avoided by
the impossibility (or, equivalently, extreme slowness) of nucleation of
the new phase, a relatively stiff equation of state may sometimes 
be maintained
over a more extended range of density than would be expected naively.
Hence, the existence of such heavy stars may not be sufficient evidence
to rule out the existence of kaon-condensation at $3-5 n_0$, especially
if these stars were born with smaller masses and
only subsequently (that is, once cold) acquired larger masses via 
accretion from companion stars.

Additionally, metastability of the sort introduced above may 
potentially be useful
in understanding the properties of GRB's or other 
poorly-understood explosive events.  A relatively light
PNS, as we have argued, may cool and deleptonize without the K-matter
mixed phase forming, even when the star's core density exceeds the
critical density for the transition.  As is evident from Fig. 
\ref{fig6}, however, the thermal kaon density increases monotonically 
and steeply with density, so that, even at the very low temperatures
$T \ll 1$ MeV eventually attained in the neutron star, there is some
density at which seeding may become possible in a reasonable time.
At very least, with increasing density, one eventually encounters the
second-order point at which the kaon field may be produced smoothly
with no need for seeding.  Hence, if an initially metastable neutron 
star begins to accrete matter from a companion binary (or, additionally,
if an initially rotating metastable neutron star gets spun down via
accretion) the central density may increase sufficiently for K-matter
to begin to appear.  

Once the kaon matter appears, however, there will be a feedback effect,
due to the softening of the equation of state.  The production of a
small quantity of K-matter in the core would allow the star to contract
slightly, thus increasing the density in the core, and increasing the
size of the region in which kaonic matter can appear.  Further kaon
production
leads to further collapse, and vice versa.  Thus if mass accretion 
and/or spin down results in the critical density for the onset of a
second-order kaon-condensate transition being reached in the neutron
star core (or, equivalently, if one reaches the low-temperature 
effective critical density at which a mixed phase can be nucleated
spontaneously with 
sufficient speed) one would expect an explosive event in which the
star contracts significantly, resulting in the release of a tremendous
amount of energy (of order tenths of $M_{sun}$).  
Cheng and Dai have discussed a similar proposal
in which accretion-induced conversion to strange quark matter is
suggested as a possible explanation for Gamma-Ray Bursters. 
\cite{chengdai}

One especially interesting aspect of such a collapse is its potentially
turbulent nature.  The picture is of a pure K-matter core seeding the 
mixed phase through several kilometers of
material above it.  In effect, the K-matter boils off of the
outer edge of the second-order core and floats upward to form the
mixed phase throughout the entire region of the mixed phase's energetic
favorability.  This implies an upward and downward transfer
of matter that closely resembles turbulent convection, but in which
strangeness rather than heat is the substance being convected.

As mentioned at the end of the previous section, more reliable 
calculations need to be performed in order to better understand
how the slowness of the weak interactions affect the original
nucleation rate estimates based on Langer's formula.  We have
argued that the correct framework for these future calculations
involves the formalism of quantum tunneling (or thermal activation)
in a theory with a (nearly) conserved global charge representing 
strangeness.  It is also worth 
mentioning that the scenario outlined here is the most 
realistic potential application of the Q-ball nucleation formalism
developed in \cite{lee1,lee2};
theorists up to now have relied on supersymmetric models to find
possible theories containing charged scalars supporting 
non-topological solitons.

\begin{center}

\large{ \textbf{Acknowledgements}}

\end{center}
Sanjay Reddy, Eduardo Fraga, and Guy Moore are acknowledged for
helpful discussions, though any inaccuracies in the present paper
are purely the responsibility of the author.  This work is supported
in part by a National Science Foundation Graduate Research Fellowship
and by the US Department of Energy grant DE-FG03-00ER41132.

%  references
\newcommand{\IJMPA}[3]{{ Int.~J.~Mod.~Phys.} {\bf A#1}, (#2) #3}
\newcommand{\JPG}[3]{{ J.~Phys. G} {\bf {#1}}, (#2) #3}
\newcommand{\AP}[3]{{ Ann.~Phys. (NY)} {\bf {#1}}, (#2) #3}
\newcommand{\NPA}[3]{{ Nucl.~Phys.} {\bf A{#1}}, (#2) #3 }
\newcommand{\NPB}[3]{{ Nucl.~Phys.} {\bf B{#1}}, (#2)  #3 }
\newcommand{\PLB}[3]{{ Phys.~Lett.} {\bf {#1}B}, (#2) #3 }
\newcommand{\PRv}[3]{{ Phys.~Rev.} {\bf {#1}}, (#2) #3}
\newcommand{\PRC}[3]{{ Phys.~Rev. C} {\bf {#1}}, (#2) #3}
\newcommand{\PRD}[3]{{ Phys.~Rev. D} {\bf {#1}}, (#2) #3}
\newcommand{\PRL}[3]{{ Phys.~Rev.~Lett.} {\bf {#1}}, (#2) #3}
\newcommand{\PR}[3]{{ Phys.~Rep.} {\bf {#1}}, (#2) #3}
\newcommand{\ZPC}[3]{{ Z.~Phys. C} {\bf {#1}}, (#2) #3}
\newcommand{\ZPA}[3]{{ Z.~Phys. A} {\bf {#1}}, (#2) #3}
\newcommand{\JCP}[3]{{ J.~Comput.~Phys.} {\bf {#1}}, (#2) #3}
\newcommand{\HIP}[3]{{ Heavy Ion Physics} {\bf {#1}}, (#2) #3}
\newcommand{\RMP}[3]{{ Rev. Mod. Phys.} {\bf {#1}}, (#2) #3}
\newcommand{\APJ}[3]{{Astrophys. Jl.} {\bf {#1}}, (#2) #3}


\begin{thebibliography}{99}

\bibitem{cp}
J.C. Collins and M.J. Perry
\PRL{34}{1975}{1353}

\bibitem{alcock}
C. Alcock, E. Farhi, A. Olinto
\APJ{310}{1986}{261-72}

\bibitem{fuller}
N.A. Gentile, M.B. Aufderheide, G.J. Mathews, F.D. Swesty, G.M. Fuller
\APJ{414}{1993}{701-11}

\bibitem{KN}
D. Kaplan and A. Nelson \PLB{175}{1986}{57}

\bibitem{G1}
N. K. Glendenning 
\PRD{46}{1992}{1274}

\bibitem{langer}
J.S. Langer 
\AP{54}{258-275}{1969}

\bibitem{linde}
A.D. Linde 
\NPB{216}{1983}{421-445}

\bibitem{fraga}
S. Alamoudi {\it et al}
\PRD{60}{1999}{125003}

\bibitem{olesen}
M.L. Olesen and J. Madsen
\PRD{49}{1994}{2698-2702}

\bibitem{heisel}
H. Heiselberg
hep-ph/9501374

\bibitem{iida}
K. Iida and K. Sato
Prog. Theor. Phys. {\bf 98} (1997) 277-282

\bibitem{shukla}
P. Shukla, A.K. Mohanty, and S.K. Gupta
\PRC{62}{2000}{054904}

\bibitem{GS}
N. K. Glendenning, and J. Schaffner-Bielich,
\PRL{81}{1998}{4564} \\
N. K. Glendenning, and  J. Schaffner-Bielich,
\PRC{60}{1999}{025803}

\bibitem{SW}
B. Serot and J.D. Walecka, {\it Advances in Nucl. Physics}, {\bf 16},
edited by J.W. Negele and E. Vogt (Plenum, New York, 1986)

\bibitem{Gbook}
N. K. Glendenning, {\it Compact Stars: Nuclear Physics, Particle Physics \\
and General Relativity }, (Springer-Verlag, New York, 1997)

\bibitem{pons}
J.A. Pons, S. Reddy, P. Ellis, M. Prakash, and J.M. Lattimer
\PRC{62}{2000}{034803}

\bibitem{kpe}
R. Knorren, M. Prakash, P.J. Ellis
\PRC{52}{1995}{3470-82}

\bibitem{prakash}
M. Prakash {\it et al.} 
\PR{280}{1997}{1}

\bibitem{FGB}
E. Friedman, A. Gal, and C.J. Batty 
\NPA{579}{1994}{578}\\
A. Cieply, E. Friedman, A. Gal, J. Mares
nucl-th/0104087

\bibitem{RO}
A. Ramos and E. Oset
\NPA{671}{2000}{481}

\bibitem{Gsurf}
M. Christiansen, N.K. Glendenning, and J. Schaffner-Bielich
\PRC{62}{2000}{025804}

\bibitem{pmpl}
J.A. Pons, J.A. Miralles, M. Prakash, and J.M. Lattimer
\APJ{553}{2001}{382-393}

\bibitem{norsen}
T. Norsen and S. Reddy
\PRC{63}{2001}{065804}

\bibitem{Gfinite}
M.B. Christiansen and N.K. Glendenning
\PRC{56}{1997}{2858}

\bibitem{lamb}
D.Q. Lamb {\it et al.}
\NPA{411}{1983}{449-473}

\bibitem{RPW}
D.G. Ravenhall, C.J. Pethick, and J.R. Wilson
\PRL{50}{1983}{2066-9}

\bibitem{HPS}
H. Heiselberg, C.J. Pethick, and E.F. Staubo
\PRL{70}{1993}{1355-9}

\bibitem{Loren}
J. Lorenzana, C. Castellani, and C. DiCastro
cond-mat/0010092

\bibitem{muto}
T. Muto, T. Tatsumi, and N. Iwamoto
\PRD{61}{2000}{063001} \\
T. Muto, T. Tatsumi, and N. Iwamoto
\PRD{61}{2000}{083002}

\bibitem{dai}
Z. Dai, T. Lu, and Q. Peng
\PLB{319}{1993}{199-202}

\bibitem{ghosh}
S.K. Ghosh, S.C. Phatak, and P.K. Sahu
\NPA{596}{1996}{670-83}

\bibitem{griest}
K. Griest, E.W. Kolb, and A. Massarotti
\PRD{40}{1989}{3529}

\bibitem{metaxas}
D. Metaxas
\PRD{63}{2001}{083507}

\bibitem{coleman}
S. Coleman
\NPB{262}{1985}{263}

\bibitem{benson}
K.M. Benson and L.M. Widrow
\NPB{353}{1991}{187}

\bibitem{lee1}
K. Lee
\PRL{61}{1988}{263}

\bibitem{lee2}
K. Lee
\PRD{50}{1994}{5333}

\bibitem{zhang}
W. Zhang, T.E. Strohmayer, and J.H. Swank
\APJ{482}{1997}{L167}

\bibitem{kerk}
M.H. van Kerkwijk
astro-ph/0001077

\bibitem{orosz}
Jerome A. Orosz and Erik Kuulkers
astro-ph/9901177

\bibitem{chengdai}
K.S. Cheng and Z.G. Dai
\PRL{77}{1996}1210

\bibitem{landau}
E.M. Lifshitz and L.P. Pitaevskii {\it Physical Kinetics} 
(Pergamon Press, 1981)

\end{thebibliography}
\end{document}